\documentclass{aa}
\usepackage{txfonts}
\usepackage{graphicx}
\usepackage{natbib}
\usepackage{amssymb}
\bibpunct{(}{)}{;}{a}{}{,}

\newcommand{\sdss}{\emph{SDSS}}

\newcommand{\teff}{$T_{\rm eff}$}
\newcommand{\logg}{$\log{g}$}

\newcommand{\masy}{${\rm mas\,yr^{-1}}$}
\newcommand{\msun}{$\rm{M_\odot}$}
\newcommand{\kms}{$\rm{km~s^{-1}}$}
\newcommand{\masyr}{$\rm{mas~yr^{-1}}$}

\newcommand{\M}{$[M/H]$}

\newcommand{\pmp}{$\pm$}

\begin{document}


\hyphenation{ana-lysed do-mi-nant}
\title{A halo blue straggler on a highly eccentric retrograde orbit
\thanks{Based on data from the Sloan Digital Sky Survey}
\fnmsep\thanks{Based on observations at the 
3.5\,m telescope at DSAZ observatory (Calar Alto) in Spain. Program ID:
H09-3.5-008}}
\author{A. Tillich\inst{1}\and N. Przybilla\inst{1}\and R.-D. Scholz\inst{2}
\and U.  Heber\inst{1}
}
\offprints{A.~Tillich
\email{Alfred.Tillich@sternwarte.uni-erlangen.de}}
\institute{Dr. Remeis-Sternwarte Bamberg\ \& ECAP, Astronomisches Institut, Friedrich-Alexander-Universit\"at Erlangen-N\"urnberg, 
Sternwartstr.7, D-96049 Bamberg, Germany\and Astrophysikalisches Institut Potsdam, An der Sternwarte 16, 
D-14482 Potsdam}

\date{Received / Accepted }

\abstract{
Blue straggler, which are stars that appear to be younger than they should be, are an important population of unusual stars in both stellar clusters and the halo field of the Galaxy. Most formation scenarios evoke either stellar collisions or binary stars that transfer mass or merge.
}
{
We investigate high-velocity stars in the Galactic halo and perform a spectral and kinematical 
analysis to shed light on their nature and origin. 
Here we report that SDSSJ130005.62+042201.6 (J1300+0422 for short) is 
an A-type star of unusually large radial velocity (504.6 $\pm$ 5 \kms). 
}
{
From a quantitative NLTE (and LTE) spectral analysis of medium-resolution optical spectra, the elemental composition is derived. Proper motion measurements combined with a spectroscopic distance estimate allow us to determine its present space velocity. Its kinematical properties are derived by integrating the equation of motion in the Galactic potential. 
}
{
We find J1300+0422 to be metal poor ([M/H]=$-1.2$) and exhibit an $\alpha$-element enrichment 
($0.3-0.4$~dex) that is characteristic of the halo population, as confirmed by a kinematical analysis of its 3D space motions, which places it on a highly eccentric retrograde Galactic orbit.
}
{
The mass of J1300+0422 (1.15 $\pm$ 0.10 M$_\odot$) is higher than the globular cluster turn-off masses indicating that it is a halo blue straggler star.
At a Galactic rest-frame velocity
of $\approx$467~\kms, the star travels faster than any known blue straggler but is still bound to the Galaxy.
}



\maketitle

\section{Introduction}\label{sec:intro}

Blue straggler stars (BSS) were first discovered as an unusual subclass of stars in globular clusters \citep{1953AJ.....58...61S}. They lie 
 on or near the main sequence but are more luminous than the turn-off stars indicating that they are of higher mass than the latter. 
Apparently, BSS are present in all the Galactic globular clusters \citep{2003ASPC..296..263P}. 

 Because the stars in a cluster are believed all to have formed at the same time, whereas the stellar 
turnoff age decreases with mass, 
 BSs should have evolved off the main sequence to become giants and white dwarfs long ago. It is generally believed that blue stragglers are coeval with the remaining stars in the cluster and originate in normal main-sequence stars that gained mass by means of a recent accretion episode. 
Most formation scenarios evoke mass transfer in and/or the merger of binary stars or collisions of stars.


It remains unclear whether two types of BSS exist in clusters, which may result from the different suggested mechanisms discussed frequently in the 
literature. The discovery of two distinct sequences of BSS in the globular cluster M~30 \citep{2009Natur.462..1128N} indicates that indeed both formation mechanisms are at work.

In open clusters, this does not seem to be the case, because \citet{2009Natur.462..1132N} indentified a high frequency of binaries among the blue stragglers in the open cluster of NGC~188, most having eccentric orbits with periods of about 1000 days. \citet{2009ApJ...690.1639B} argue that most if not all BSS in open clusters arise from multiple star systems. This suggests that blue stragglers
are formed in both ways in star clusters,
with collisions/mergers becoming more common with
increasing cluster density.


In the Galactic field, it is more difficult to identify BSS because the stellar age cannot be determined. Field blue stragglers are therefore usually identified as metal-poor main sequence objects that are hotter than globular cluster main-sequence turnoff stars. 
Nevertheless, many blue metal-poor stars $[Fe/H]<-1$ with main-sequence luminosities have been found to be hotter than the main-sequence turnoff of globular clusters and are therefore considered to be field analogs of the cluster BSS population. These metal-poor stars seem to be so numerous that their specific frequency of appearance relative to regular horizontal branch stars was found to be higher than in globular clusters \citep{2000AJ....120.1014P}. Since the stellar density is low in the 
field, binary star evolution and mass transfer is probably the most common path of formation among 
field BSS. 
\cite{2001AJ....122.3419C,2005AJ....129.1886C} pointed out that some BS candidates among metal-poor halo main-sequence stars are not binaries, 
hence binary star evolution
and mass transfer may not be the only path.
They argue that the apparently single stars could be the partial remnants of an accreted
dwarf satellite galaxy whose star formation continued
over a long period of time. These metal-poor main-sequence stars are simply therfore young stars.


A detailed spectral analysis of a blue metal-poor star sample in the field by \cite{2000AJ....120.1014P} measured a 
very high binary fraction of at least 67\%, dominated by long period (wide) binaries. This population 
appears to be very similar to that of the open 
cluster NGC~188 \citep{2009Natur.462..1132N}.  
Assuming a formation by Roche lobe overflow during the red giant branch stage 
of the primary, \cite{2000AJ....120.1014P} identified at least half of the 
blue metal-poor stars in their survey as blue stragglers.
This is supported by more recent and much larger surveys for faint blue stars in the halo. 
For example, \cite{2008ApJ...684.1143X} found that blue stragglers account for 
half of their sample of more than 10\,000 A-type stars. 


We have embarked on a search for so-called hyper-velocity stars \citep[HVS,][]{2009A&A...507L..37T} based on the work of \citet{2008ApJ...684.1143X}, who presented radial velocities for a large sample of blue stars. 
Their sample is a mix of blue horizontal
branch (BHB), blue straggler, and main-sequence stars with effective temperatures
roughly between 7000 and 10\,000\,K according to their colours.

To study the stellar motions in 3D, we need to derive their tangential velocities from proper motions and spectroscopic distances.
The measurement of proper motions for faint high-velocity stars 
is the most challenging part of this project because the stars' distances are large and therefore 
require highly accurate proper motion measurements.

We focused on the fastest stars in terms of large Galactic 
rest-frame (GRF) velocities with the aim of determining their nature, distance, and
kinematics from detailed quantitative spectral analyses and astrometry. 
First results have already been reported \citep{2009A&A...507L..37T,2009_J1539}. 
We found J0136+2425 to be an A-type 
main-sequence star travelling at $\approx$590~\kms, possibly unbound to the 
Galaxy \citep{2009A&A...507L..37T}, which makes it an excellent HVS candidate.
More importantly, it was inferred to have originated in the outer Galactic rim, nowhere near the Galactic centre, which would be the favoured place of origin if a supermassive black hole acts as a slingshot as suggested by \citet[][see also \cite{2009ApJ...690.1639B}]{1988Natur.331..687H}.
Amongst the stars of negative 
GRF velocity, we discovered that J1539+0239 is a BHB travelling with the largest space velocity of any BHB star known so far, which allowed us to 
place a lower limit on the mass of the Galactic halo \citep{2009_J1539}. 

Here we report that 
SDSSJ130005.62+042201.6 (J1300+0422 for short) is a metal-poor blue straggler of Population II 
on a wide retrograde halo orbit that is only marginally bound to the Galaxy in a standard Galactic 
potential.




\section{Target selection and proper motions}\label{sec:tar+pm}

\begin{figure}[t]
\begin{center}
\includegraphics[scale=0.4]{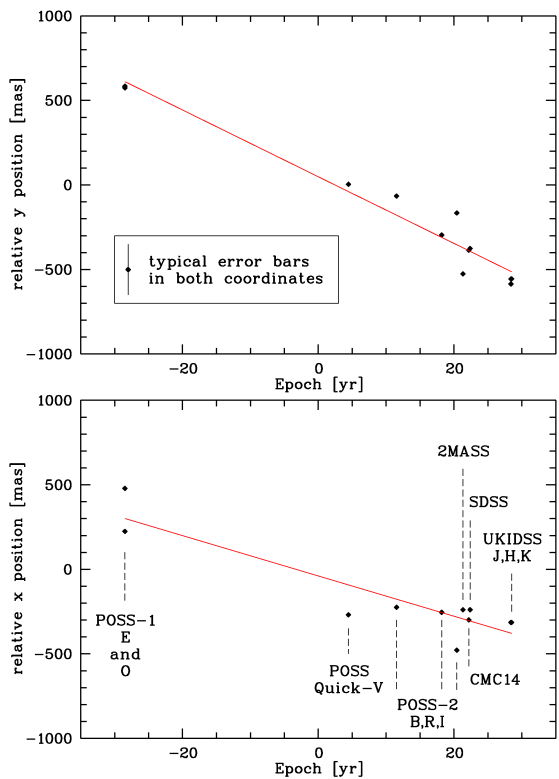}
\caption{\label{fig_PMfit}Linear fit of the position measurements for J1300+0422, where 1978.84 is the zero epoch.}
\end{center}
\end{figure}

We selected all stars 
with GRF velocities $v_{\rm GRF}>+350$~\kms\ from the 
RV-based sample of \cite{2008ApJ...684.1143X} and obtaining 11 
targets for which we attempted to measure proper motions \citep{2009A&A...507L..37T}. 
All 
available independent position measurements from Schmidt plates
(APM - \cite{2000yCat.1267....0M};
SSS - \cite{2001MNRAS.326.1279H}) were collected
and combined with the \sdss\ and other available positions
(CMC14~\cite{2006yCat.1304....0C};
2MASS - \cite{2003tmc..book.....C};
UKIDSS - \cite{2007MNRAS.379.1599L})
to perform a first linear proper-motion fit. However, there were even
more measurements from Schmidt plates, for up to 14 different epochs in the case
of overlapping plates of the Digitised Sky
Surveys\footnote{http://archive.stsci.edu/cgi-bin/dss\_plate\_finder}.
FITS images of 15 by 15 arcmin size were extracted from all available plates
 and ESO MIDAS tools were used to measure positions. For this purpose,
we selected compact background galaxies around each target, identified from
\sdss, to transform the target positions on all the Schmidt plates to the
\sdss\ system. The small fields allowed us to apply a simple model
(shift+rotation) and to achieve a higher accuracy in our final proper motion fit 
for all our targets (see e.g. Fig.~\ref{fig_PMfit} and Fig.~\ref{fig_PMfit1553}). 
Proper motions differing significantly from zero were found for the three brightest stars only\footnote{SDSSJ155352.41+003012.0 (J1553+0030 for short) turned out to be a spectroscopic binary and will not be studied further in this paper. For the sake of completeness, we provide details in the Appendix}. 
The results for J1300+0422 are given in Table~\ref{tab_HVS}.
 
\section{Observations and quantitative spectroscopy}\label{sec:spec}

To search for radial velocity variations, we reobserved J1300+0422 with 
the TWIN spectrograph at the 3.5m telescope on Calar Alto on   
May 2009. The radial velocity (heliocentric $v_{\rm rad}$ = 515.3 $\pm$ 8 \kms) was derived by 
$\chi^2$-fitting of appropriate synthetic spectra over the full spectral range.
\footnote{Because we used many spectral lines our results differ from those of
 \citet{2008ApJ...684.1143X}, who used cross-correlation for RV and then the 
$\mathrm H{\gamma}$+$\mathrm H{\delta}$ lines for stellar-type determination only.} 
It was found to be consistent with that measured from the \sdss\ spectrum (heliocentric $v_{\rm rad}$ = 504.6 $\pm$ 5 \kms).

We also inspected three individual \sdss\ spectra that were acquired one after the other. No variations were found.  
This indicates that large radial velocity variations on small timescales are unlikely, but we cannot rule out small RV variations over long periods as observed in many BSS 
\citep{2005AJ....129.1886C}.  
 

A quantitative analysis of the coadded \sdss\ spectrum of J1300+0422 was carried
out following the hybrid NLTE approach discussed by
\cite{2006A&A...445.1099P}. In brief, line-blanketed LTE model 
atmospheres were computed with ATLAS9 \citep{1993KurCD..13.....K} and NLTE (and
LTE) line-formation calculations were performed using updated versions of 
{\sc Detail} and {\sc Surface} \citep{1981PhDT.......113G,butler_giddings_1985}. Many
astrophysical important chemical species were treated in NLTE, using
state-of-the-art model atoms (H: \citealp{2004ApJ...609.1181P}; \ion{C}{i}:
\citealp{2001A&A...379..936P}; \ion{N}{i}: \citealp{2001A&A...379..955P};
\ion{O}{i}: \citealp{2000A&A...359.1085P}; \ion{Mg}{i/ii}: \citealp{2001A&A...369.1009P}; 
\ion{Ti}{ii} and \ion{Fe}{ii}: \citealp{1998ASPC..131..137B}).

\begin{figure*}
\begin{center}
\resizebox{.99\hsize}{!}{\includegraphics{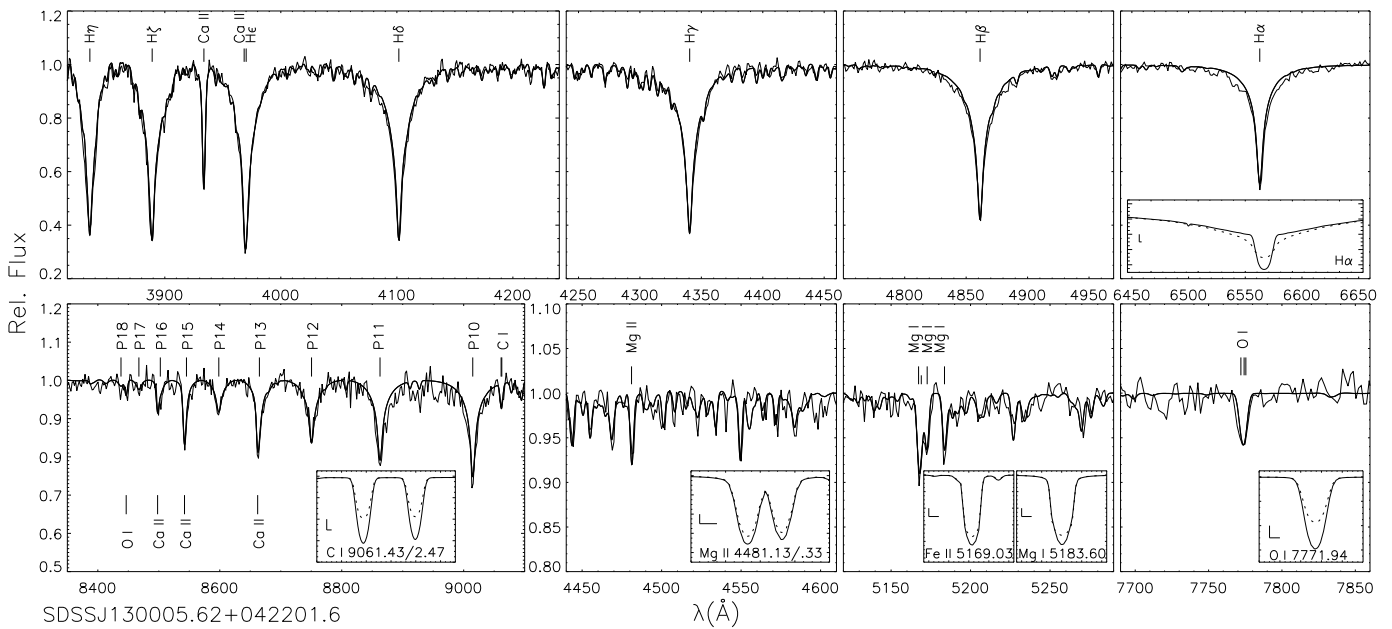}}\\[1.95mm]
\caption{\label{linefits}
Comparison of NLTE spectrum synthesis (thick line) with observation
(thin wiggly line) for J1300+0422. Displayed are selected regions around 
the Balmer lines, the higher Paschen series, \ion{Mg}{ii}\,$\lambda$4481{\AA}, 
the \ion{Mg}{i}\,b, and the near-IR \ion{O}{i} triplets. The insets
compare computed NLTE (full) and LTE profiles (dotted lines) for
several key features, as indicated. No convolution by instrumental or
rotational profiles was applied to exemplify what can be expected at
high spectral resolution. The markers are extended horizontally by 0.1\,{\AA} 
and vertically by 0.1 units.}
\end{center}
\end{figure*}

The effective temperature {\teff} and the surface gravity $\log g$ 
were determined by fits to the Stark-broadened Balmer and Paschen
lines and the ionisation equilibrium of \ion{Mg}{i/ii}. We adopted a 
microturbulent velocity of 2\,km\,s$^{-1}$ and a 
mixing-length to pressure-scale-height ratio $\ell/H$ of 1.25 for
the convective atmosphere. The stellar
metallicity was derived by model fits to the observed metal-line
spectra. Results are listed in Table~\ref{tab_HVS} and a
comparison of the resulting final synthetic spectrum with
observations is shown in Fig.~\ref{linefits}. Overall, excellent agreement 
is obtained for the strategic spectral lines throughout the entire 
wavelength range. The uncertainties in the stellar parameters were 
constrained by the quality of the match of the spectral indicators 
within the given S/N limitations.

It may be instructive at this point to take a closer look at the
spectrum synthesis in both NLTE and LTE, as this has not been done so far
for Population\,II blue straggler stars\footnote{\cite{2005ApJ...632..894D}
presented global fits of partially line-blanketed NLTE
model atmospheres to several blue straggler stars in globular
clusters. However, the differences between NLTE and LTE model spectra
were not elaborated on there.}. A few comparisons of NLTE
and LTE profiles are therefore shown in the insets in Fig.~\ref{linefits}.
The combination of a higher {\teff} than typically found for
Population\,II stars and the diminished line blocking because of the low metal
content results in a hardened radiation field that caused 
pronounced NLTE effects on many diagnostic lines. NLTE strengthening
is found for the Doppler core of H$\alpha$ -- in line with the
behaviour in cooler \citep[e.g.][]{2004ApJ...610L..61P} as well as in
hotter stars \citep[e.g.][]{2007A&A...467..295N} -- but the inner line wings are
weakened. The higher Balmer and Paschen lines show much lower
deviations from LTE. Our calculations predict that the majority of the weak
metal-lines are described well by the assumption of LTE. 
On the other hand, many of the stronger metal lines, such as e.g.,
those shown in the insets in Fig.~\ref{linefits}, exhibit pronounced NLTE
strengthening. To reproduce the NLTE equivalent widths of
these particular lines, we need to apply abundance corrections 
of about +0.2\,dex (\ion{Mg}{i/ii}, \ion{Fe}{ii}), +0.5\,dex
(\ion{O}{i}), and +0.8\,dex (\ion{C}{i}) in LTE. We note, however,
that these line are close to being saturated. LTE
computations for higher abundances cannot therefore not reproduce the
NLTE line depths at all, but instead have stronger line wings.
Abundance studies based on equivalent widths may therefore be
misleading, because these differences remain unnoticed. An investigation
at high spectral resolution would therefore be worthwhile 
to facilitate the NLTE effects to be studied in detail.

\begin{figure}
\begin{center}
\includegraphics[scale=0.7]{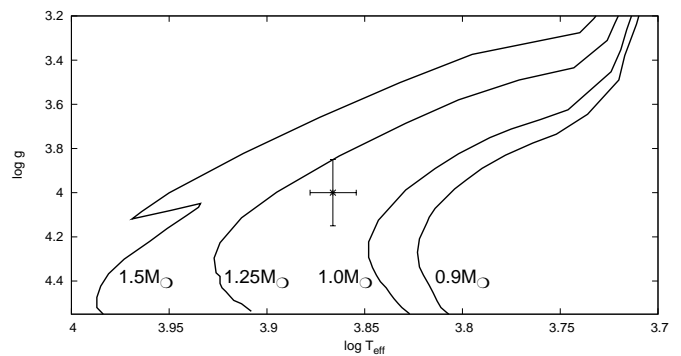}
\caption{\label{teff_g_J1300}
Comparison of the position of J1300+0422 in the (\teff,\logg) diagram to 
evolutionary tracks of \citet{1992A&AS...96..269S} for subsolar metallicity ($Z=$0.001), to 
determine its mass.
}
\end{center}
\end{figure}

Its measurement of {\teff} and gravity places J1300+0422 on the main sequence 
(see Fig.~\ref{teff_g_J1300}) at a mass of 1.15\,M$_\odot$ as
derived by comparing the position of the star to predictions of the
evolutionary models of \cite{1992A&AS...96..269S}.
No rotational broadening was detected. The metallicity is
lower than solar by a factor of almost 20 and the abundances of the 
$\alpha$-elements (those from O to Ti)
are enhanced by $\sim$0.3--0.4\,dex, which is typical of the halo population.
This suggests that star is a halo blue straggler.
All results are summarised in Table \ref{tab_HVS}.

\begin{table}
\caption{Results of the \textbf{analysis of J1300+0422.} 
The resulting GRF velocity
v$_{\rm GRF}$ and the local
escape velocity v$_{\rm esc}$, also in a more massive halo v$_{\rm esc}^{\rm a}$, are given.}
\label{tab_HVS}
\begin{center}
\begin{tabular}{lc|lc}
\hline\hline
$V$\ (mag)$^{\rm a} $     & 15.12 $\pm$ 0.02 & 
$E(B-V)$\ (mag)$^{\rm b} $ &  0.05 $\pm$ 0.02 \\
$\mu_\alpha\cos(\delta)$\,(\masyr) &   $-$11.9 $\pm$ 1.7 & 
$\mu_\delta$\,(\masyr) &   $-$19.7 $\pm$ 1.4 \\
$n_{\rm ep}$ & 12 & $n_{\rm gal}$ & 15 \\
\teff\ (K)      &  7350 $\pm$ 200  & 
 \logg (cgs)      &  4.00 $\pm$ 0.15  \\
 \M          &  $-$1.2$\alpha$        &  
 $M/M_\odot$ &  1.15 $\pm$ 0.10   \\
 $v_{\rm rad}$ (\kms)   & 504.6 $\pm$ 5 &
 $v_{\rm rot} \sin i$ (\kms) & - \\ 
 $d$\ (kpc)    & 3.25 $\pm$ 0.62  &    
  $v_{\rm GRF}$ (\kms) & 467 $^{+41}_{-21}$   \\
  $v_{\rm esc}$ (\kms) & 529 & $v_{\rm esc}^{\rm a}$ & 708    \\
 \hline  
 \multicolumn{4}{l}{\small $^{\rm a}$ The visual magnitude has been
derived from \sdss\ photometry}\\
 \multicolumn{4}{l}{following \cite{2006A&A...460..339J} }\\
 \multicolumn{4}{l}{\small $^{\rm b}$ The interstellar colour excess
$E(B-V)$\ has been determined}\\
 \multicolumn{4}{l}{\small 
 by comparing the
observed colours to synthetic ones}\\
 \multicolumn{4}{l}{\small 
 from the model spectral energy
distribution. }\\

\end{tabular}
\end{center}
\end{table}

\section{Distance, kinematics, and errors}\label{sec:kine}

Using the mass, effective temperature, gravity, and extinction-corrected apparent
magnitude we derive the distance following \cite{2001A&A...378..907R} 
using the fluxes from
the final model spectrum. The distance error is dominated by the gravity error. 

\begin{figure}
\begin{center}
\includegraphics[scale=0.65]{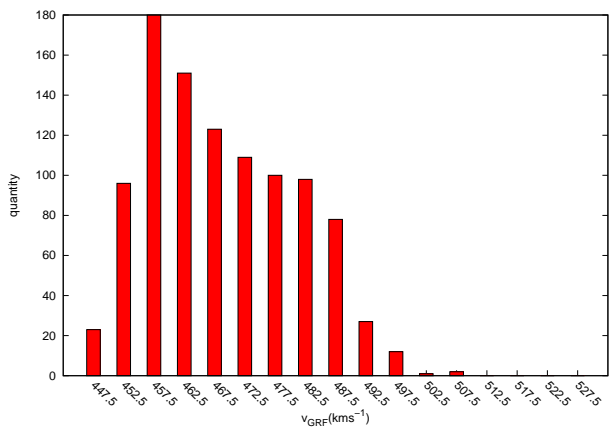}
\caption{\label{fig:veldistrib_J1300}   
Galactic rest-frame velocity distribution for J1300+0422 derived from a Monte Carlo procedure
with a depth of 1000. 
}
\end{center}
\end{figure}

Applying the Galactic potential of
\citet{1991RMxAA..22..255A}, we calculated orbits and 
reconstructed the path of the star through the Galactic halo with the program of
\citet{1992AN....313...69O}. 
The distance of the GC from the Sun was adopted to be 8.0~kpc and
the Sun's motion with respect to the local standard of rest was taken from 
\citet{1998MNRAS.298..387D}. 
Since the RV is well known, the error in the space motion is dominated
by that of the distance, which is affected most by the gravity error, and those of the 
proper
motion components. Varying these three quantities within their respective errors, 
we applied a Monte Carlo procedure to derive the distribution of GRF velocities at the present 
location and its median (see Fig.~\ref{fig:veldistrib_J1300}). 
The local escape velocity was calculated by assuming the Galactic potential
of \citet{1991RMxAA..22..255A}. 
%
The GRF velocity of 467$^{+41}_{-21}$~\kms\ 
therefore remains below the local escape velocity of $v_{\rm esc}\approx529$~\kms 
and J1300+0422 is bound to the Galaxy. 

To quantify the kinematics of the star more accurately, we compared our results with 
those of \cite{2006A&A...447..173P}. Based on the 3D-orbit, the $V-U$ diagram, the 
eccentricity \textit{e}, and the the $Z$-component of the angular momentum $J_{\rm Z}$ 
these authors introduced a kinematic population classification scheme 
and combined it with age information. They analysed 398 DA white dwarfs from the SPY 
project, the largest homogeneous sample of its kind, and performed a 
detailed kinematic analysis accounting for errors using a Monte Carlo error 
propagation code, as in our method. 
A substantial thick disk fraction of 7\% was found, while 2\% of the DAs appeared to have 
characteristic halo properties. 

\begin{figure*}
\includegraphics[scale=0.72]{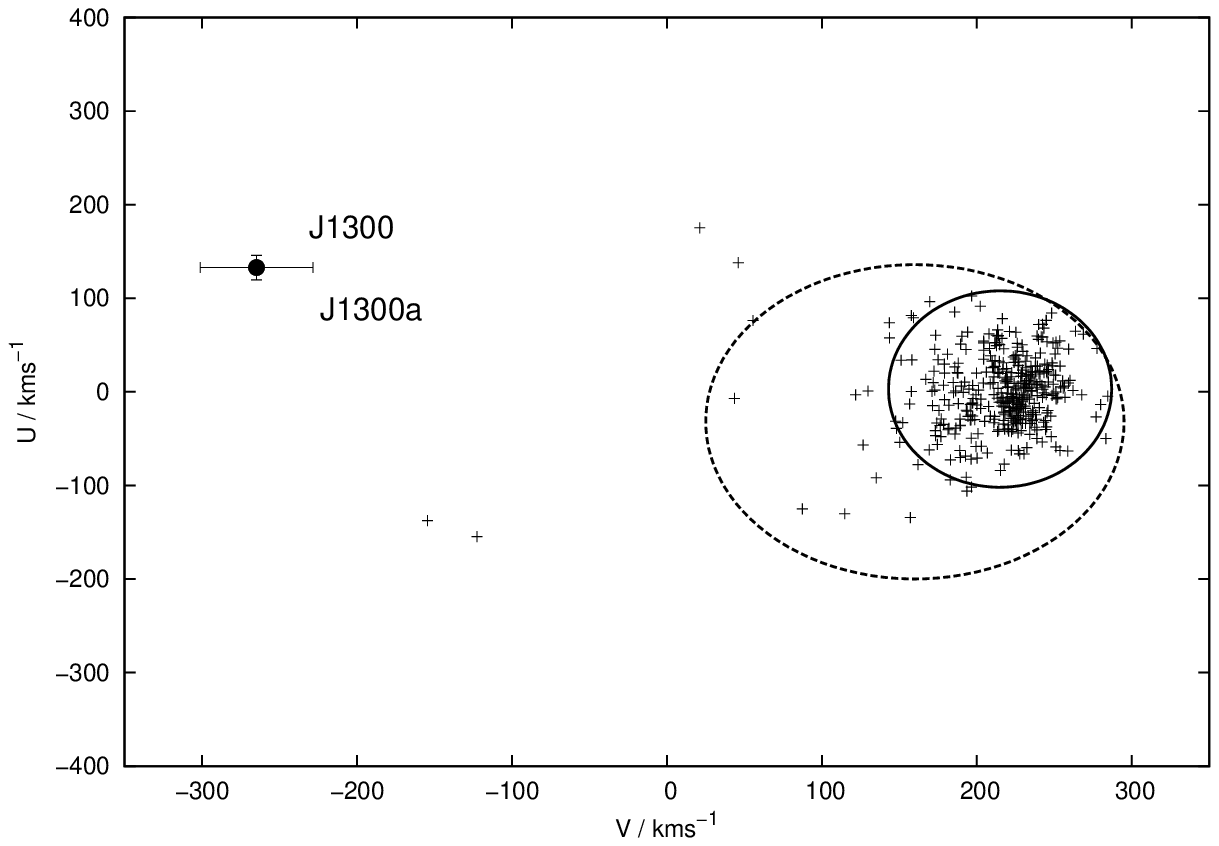}
\includegraphics[scale=0.72]{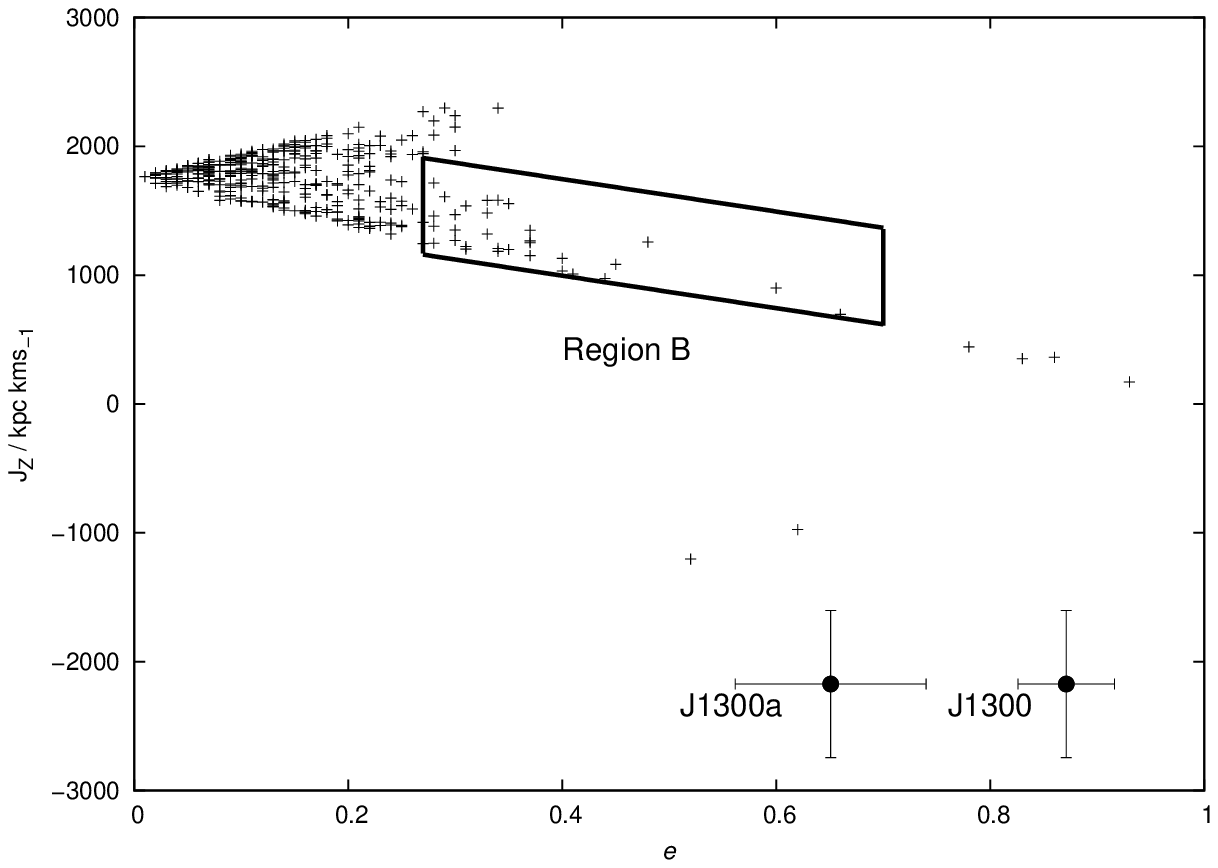}
\caption{\label{fig:kine}   
$V-U$ (\textit{left}) and \textit{e}-$J_{\rm Z}$ diagram (\textit{right}) for J1300+0422. In case of a 
higher Galactic halo mass \citep{2009_J1539} the position is marked with J1300a. The white dwarf 
sample (+) of \cite{2006A&A...447..173P} serves as reference. The ellipses in the
V$-$U diagram indicate the thin and thick disk 
contours, halo stars lie outside of these ellipses. 
The left-hand triangle-shaped cloud of stars are the thin disk white dwarf population, while the solid box marks the thick disk region (Region B). Note that in both diagrams J1300+0422 lies far away from the disk populations.
}
\end{figure*}

\begin{figure*}
\includegraphics[scale=1.3]{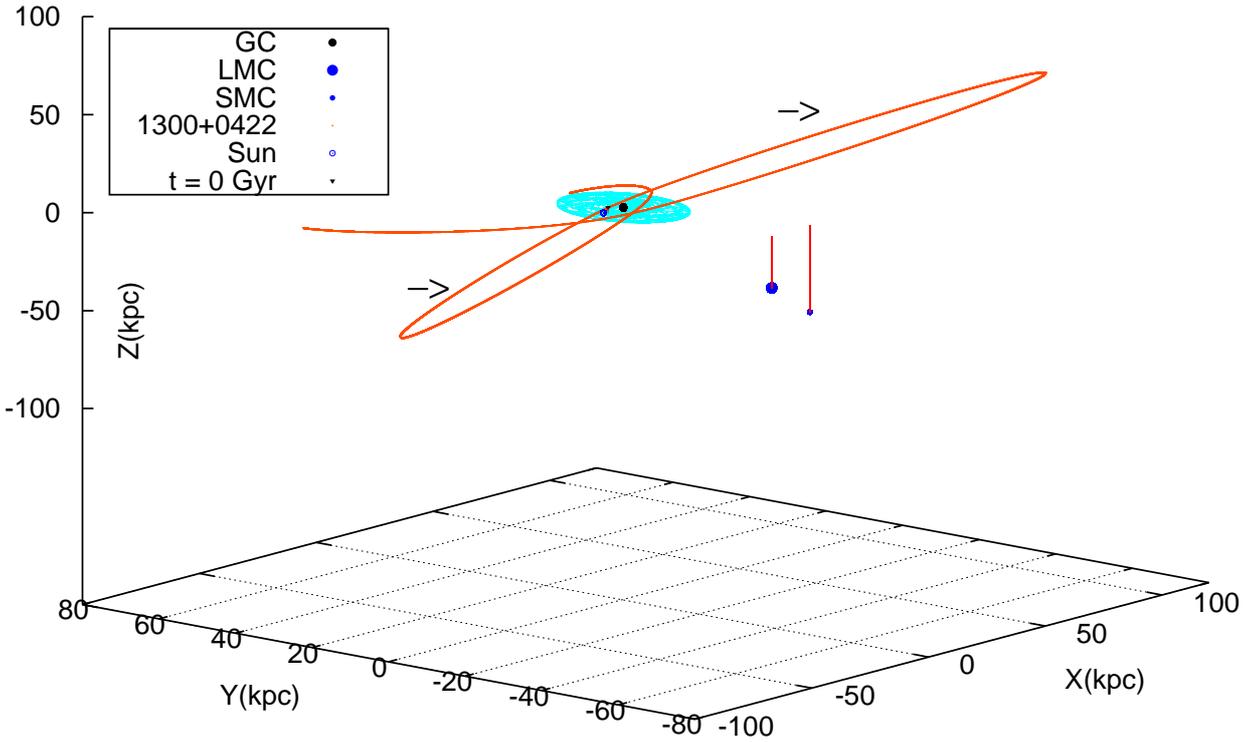}
\caption{\label{fig_J1300_3dplot}
Trajectory of the metal-poor blue straggler J1300+0422, relative
 to the Galactic disk (grey), calculated for \pmp2~Gyrs in time. The current position is marked. 
}
\end{figure*}

Following this approach, we found that J1300+0422 clearly belongs to the halo population. 
This is already obvious in Fig.~\ref{fig:kine}, which shows the characteristic 
$V-U$ and -\textit{e}-$J_{\rm Z}$ diagrams in relation to the reference white dwarf sample of 
\cite{2006A&A...447..173P}. We note that the orbit of J1300+0422 is retrograde and highly eccentric. 
Furthermore, the star travels far out into the halo out to distances of more than 
100~kpc (see Fig.~\ref{fig_J1300_3dplot}). 
Whether the star is bound to the Galaxy depends on the Galactic
potential adopted, in particular on the dark matter halo, as pointed out by 
\citet{2009ApJ...691L..63A}. In our standard potential \citep{1991RMxAA..22..255A}, we 
adopted a halo mass out to 100~kpc of $M_{\rm Halo}$= 8$\times$10$^{11}M_\odot$.  
The analysis of J1539+0239 \citep{2009_J1539} inferred a higher halo mass of at least $M_{\rm halo}\sim1.7_{-1.1}^{+2.3}\times10^{12}$\,\msun. We therefore 
repeated our analysis accordingly and found a higher escape velocity of 
$v_{\rm esc}\approx708$~\kms, which means that the star would in any case be bound. 
The effect on the angular momentum was negligible, while the orbit became significantly 
less eccentric (see Fig.~\ref{fig:kine}). Hence the star might have crossed the disk even 
more often within its lifetime.

%

\section{Conclusion}\label{sec:conclusion} 

We have presented a quantitative spectral analysis of a high-velocity star from 
the sample of faint blue stars in the halo of \citet{2008ApJ...684.1143X}.
Its radial velocity, proper motion, and spectroscopic distance were derived and
a detailed kinematical analysis was performed using the Galactic potential 
of \citet{1991RMxAA..22..255A} as well as a potential modified one that assumes 
a more massive dark matter halo. 


The metal-poor A-type star J1300+0422 was identified as a blue straggler of 
1.15~M$_\odot$\ due to its main-sequence gravity. A detailed NLTE analysis was 
performed, which we compared to the standard LTE approach. 
Significant differences were found especially for \ion{C}{i} and \ion{O}{i}. 
With its low metallicity of 
$[Fe/H]=-1.2$ and characteristic enhancement of $\alpha$-elements, it 
would fit perfectly into the sample of \cite{2000AJ....120.1014P}, apart from the 
huge space velocity of the star. The kinematic characteristics ($U$, $V$, $e$, $J_{\rm Z}$) 
confirm the halo membership of J1300+0422 beyond any doubt. 
In addition, its trajectory continues far out into the halo. 
Many blue stragglers were found to be long period binaries 
\citep[with periods of several $100$\ to $1000$~days;][]{2005AJ....129.1886C,2009Natur.462..1132N} with low 
radial velocity semi-amplitudes ($K\sim 5-10$~\kms, \citealp{2001AJ....122.3419C,2005AJ....129.1886C} ). Whether or 
not J1300+0422 is such a 
binary needs to be verified by an extensive radial velocity study. 

Our kinematical result is limited by the errors in both the spectroscopic distance and the 
proper motion. ESA's upcoming astrometry mission GAIA will improve the situation because it will 
provide a parallax measurement with which to check the spectroscopic distance and improve the proper motion 
of J1300. GAIA will also have an enormous impact on research into blue stragglers in a more general 
sense as it will provide astrometry of thousands of halo BSS.

\acknowledgements{A.T. acknowledges funding by the Deutsche Forschungsgemeinschaft through  grant HE1356/45-1. Travel to the Deutsch Spanisches Astronomisches Zentrum (DSAZ, Calar Alto, Spain) was funded by DFG through grant He1356/50-1. 
We are very grateful to Stephan Geier for stimulating discussions and advice.
Our thanks go to Sebstian M\"{u}ller for observing and reducing the 
data from DSAZ. 
Funding for the \sdss and \sdss-II has been provided by the Alfred P.
Sloan Foundation, the Participating Institutions, the National Science
Foundation, the U.S. Department of Energy, the National Aeronautics
and Space Administration, the Japanese Monbukagakusho, the Max Planck
Society, and the Higher Education Funding Council for England. The
\sdss Web Site is http://www.sdss.org/.
}

\bibliography{references}
\bibliographystyle{aa}

\appendix
\section{J1553}
In the spectrum of J1553+0030, the Ca~II IR triplet and to a lesser ecxtent the 
Mg~I b triplet are unusually strong, and most likely produced by a cool companion contributing to the spectra, hence we had to exclude J1553+0030 from the rest of our analysis. 
To confirm this result, we reobserved 
J1553+0030 with the TWIN spectrograph at the 3.5m telescope on Calar Alto in  
May 2009. 

\begin{figure}[t]
\begin{center}
\includegraphics[scale=0.4]{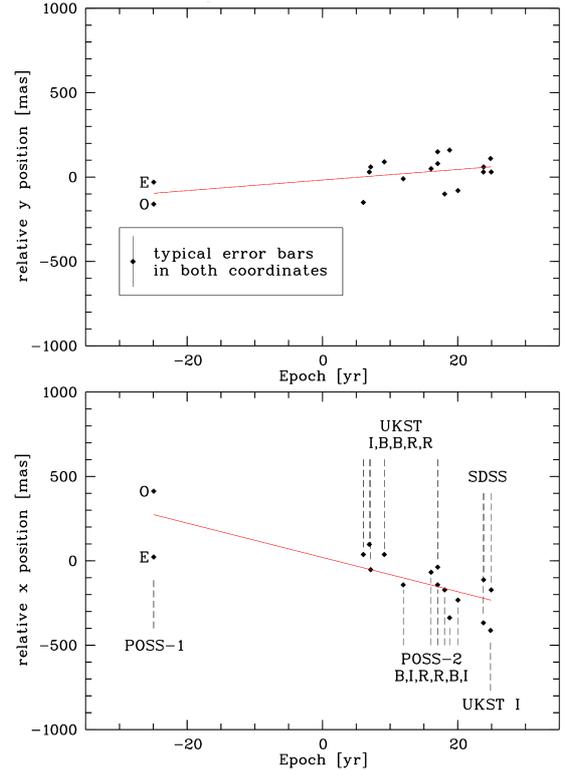}
\caption{\label{fig_PMfit1553}Linear fit to the position measurement for J1553+0030, where 1975.40 is the zero epoch. The PM was derived to be\ $\mu_\alpha\cos(\delta) = -$11.9 $\pm$ 1.7 \masy\ and\ 
$\mu_\delta =-$19.7 $\pm$ 1.4 \masy.}
\end{center}
\end{figure}


\end{document}